\documentclass[10pt,twocolumn,prl,aps,amssymb,amsmath,tightenlines]{revtex4}

\usepackage[caption = false]{subfig}
\usepackage{graphicx}
\usepackage{dsfont} 
\usepackage{bbm} 
\usepackage{verbatim}
\usepackage[normalem]{ulem}
\usepackage{enumerate}
\usepackage{mathrsfs}

\newcommand{\re}{\mbox{$\rm e$}}

\newcommand{\qed}{\hfill \ensuremath{\Box}}

\usepackage{xcolor}
\usepackage{bm}
\usepackage{txfonts}
\usepackage{amsfonts}
\usepackage{braket}
\begin{document}

\title{
Decoherence Implies Information Gain}
\author{Dorje C. Brody$^{1,2}$ and Lane P. Hughston$^{3}$}
\affiliation{
$^1$School of Mathematics and Physics, University of Surrey, 
Guildford GU2 7XH, UK \\ 
$^{2}$Institute of Industrial Science, The University of Tokyo, Tokyo 153-0041, Japan \\
$^{3}$Department of Computing, Goldsmiths University of London, New Cross,
London SE14 6NW, UK} 

\vspace{0.2cm} 
\date{\today}

\begin{abstract}
\noindent 
It is shown that if the wave function of a quantum system undergoes an arbitrary random transformation such that the diagonal elements of the density matrix in the decoherence basis associated with a preferred observable remain constant, then  (i) the off-diagonal elements of the density matrix become smaller in magnitude, and (ii) the state of the system \textit{gains} information about the preferred observable from its environment in the sense that the uncertainty of the observable is reduced in the transformed state. These results do not depend on the details of how the system-environment interaction generates the random state transformation, and together imply that decoherence leads in general to information gain, not information loss. 
\end{abstract}

\maketitle


\noindent The phenomenon of {\em decoherence} -- which we define to mean 
the reduction of the magnitudes of the off-diagonal 
elements of the density matrix of a quantum system in the basis of some preferred 
observable -- is undoubtedly one of the most important conceptual developments in 
the foundations of quantum mechanics over the past forty years. Although the debate 
continues over whether decoherence in itself resolves the measurement problem 
\cite{Adler2003,Ballentine}, it does clearly provide deep insights into the nature of the 
transformations of the state of a quantum system that can occur when a system 
interacts with its environment \cite{JZ,Schlosshauer 2010,Schlosshauer 2019,Zurek22}. It is widely held that 
decoherence represents a loss of information from a system to its environment 
\cite{Zurek1991, Zurek2009}. Here we challenge this view and show, in fact, that 
decoherence is about 
information flowing {\em into} the system {\em from} the environment, not the other 
way around. Our results are general, applying independently of the precise nature 
of system-environment interactions. 

The paper is organized as follows. First, we examine 
the traditional view that decoherence means information loss. Although this widely held view is based on what may appear to be a compelling line of reasoning, we point out that there are flaws in the traditional argument, which when corrected lead one to the opposite conclusion. 
The usual view is that 
since the von Neumann entropy of the system increases as the system decoheres, this means that information is lost from the system. 
The problem is that the von Neumann entropy does not involve the preferred basis along which the decoherence takes place. 
We show instead that {\em if the Shannon information associated with the expansion  of a quantum state in the basis of the preferred observable increases, then the density 
matrix of the system necessarily decoheres in the basis of that observable}. This result is followed 
by a more general and perhaps 
surprising conclusion -- namely, that any random transformation 
of the wave function that preserves the diagonal elements of the density matrix 
in the basis of a preferred observable gives rise to decoherence and that this 
implies the state incurs a gain in Shannon information about that observable. 

Let us begin by summarizing the argument leading to the ``information-loss'' 
interpretation of decoherence. The notion that all physical systems are ultimately open systems, facing an environment with which they interact, can be regarded as a fundamental principle of nature \cite{JZ1985, BP}. With this in mind, we consider an ideal quantum system that is initially isolated and in a  
pure state $|\psi_0\rangle$ that can be written in the form 
\begin{eqnarray}
|\psi_0\rangle = \sum_{k=1}^n \sqrt{p_k}\, \re^{{\rm i}\theta_k} \, |X_k\rangle  
\end{eqnarray} 
when expanded in terms of a set of normalized eigenstates 
$\{|X_k\rangle \}_{k = 1,\,.\,.\,.\,,\,n}$ of a certain nondegenerate observable ${\hat X}$. 

We are concerned in what follows with the overall magnitudes 
of the matrix elements of the initial density matrix  
\begin{eqnarray}
{\hat r}=|\psi_0\rangle\langle\psi_0|.
\end{eqnarray} 
The matrix elements of ${\hat r}$ in the $\hat X$-basis are given by 
\begin{eqnarray} \label{initial matrix elements}
r_{ij} =   \langle X_i  |\psi_0 \rangle  \langle  \psi_0  |X_j\rangle = 
\sqrt{p_i  p_j} \, \re^{{\rm i}(\theta_i-\theta_j)} 
\end{eqnarray} 
for $i, j = 1,\ldots,n$, whereas for the overall magnitudes one has 
\begin{eqnarray}\label{magnitudes for initial state}
|\,r_{ij}\,| =  \left| \langle X_i  |\psi_0 \rangle  \langle  \psi_0  |X_j\rangle\right| = 
\sqrt{p_i  p_j} \,.
\end{eqnarray} 

Then we place the system in an environment  
with many more degrees 
of freedom than the system and we allow the system to interact with it.  This interaction will lead to entanglement of the system with the environment and from there to decoherence of the reduced density matrix in the frame of a preferred basis. We account for the effects of entanglement with the environment by assuming that the interaction with the environment perturbs the system in such a way as to generate an effective state vector transformation of the form
\begin{eqnarray} 
|\psi_0\rangle \to |\Psi\rangle = \sum_{k = 1}^n \sqrt{\pi_k}\, \re^{{\rm i}\phi_k} |X_k\rangle, 
\label{random transformation}
\end{eqnarray} 
in the preferred basis, where $\{\pi_i\}$ and $\{\phi_i\}$ are modelled as {\em random variables}. This step is justified if one notes that an observer will typically have no access to detailed information about the 
perturbation and hence can at best merely speak about the statistics of the perturbed 
state. This leads us to a ``reduced form" model for the effects of decoherence. 

It should be emphasized that rather than modelling decoherence by introducing a linear phase-damping channel acting on the density matrix of the system, as discussed, for example, in \cite{HZ2012}, our reduced form method introduces a generic non-linear map on the space of density functions, induced by a random transformation of the underlying pure state, of the form \eqref{random transformation}. 

The form that this map takes depends on specific characteristics of the environment, though as we shall see later the unifying feature of such transformations is that the probabilities associated with the various elements of the decoherence basis satisfy a {\em stochastic conservation law}, that is to say, they have the ``martingale" property. As concrete examples of such nonlinear random pure state maps we refer to the well known energy-based dynamical models for quantum state reduction (see, e.g., \cite{Gisin1989}, \cite{BH2023}, and references cited in the latter), which indeed explicitly exhibit decoherence \cite{ABBH2001}. 

Since the expectation value of any observable ${\hat F}$ in a random pure state 
$|\Psi\rangle$ takes the form ${\rm tr}\,[{\hat F}\,|\Psi\rangle\langle\Psi|]$, it follows that 
the state of the system, insofar as an external observer is concerned, is given by the overall mean density 
matrix
\begin{eqnarray} \label{transformed density matrix}
{\hat\rho}={\mathbb E}\,[\,|\Psi \rangle\langle\Psi |\,], 
\end{eqnarray} 
where ${\mathbb E}[-]$ 
denotes the ensemble average (i.e.~the expectation) over the random degrees of freedom. Thus, the random state vector $|\Psi\rangle$ 
represents an ``unravelling" of $\hat\rho$  \cite{Carmichael, WD}. 

We are interested in determining conditions on the random probabilities and phases in the transformation \eqref{random transformation} sufficient to ensure that density matrix \eqref {transformed density matrix} exhibits decoherence. We do not insist on complete decoherence in the basis of the preferred observable, merely that interaction with the environment induces some degree of decoherence, which may or may not amount to a complete dephasing in the decoherence basis. The matrix elements of ${\hat\rho}$, in the ${\hat X}$-basis, are 
\begin{eqnarray} \label{matrix elements of state}
\rho_{ij} =  {\mathbb E}\left[  \langle X_i  |\Psi \rangle  \langle  \Psi  |X_j \rangle \right]= {\mathbb E}\left[ \! \sqrt{\pi_i \,\pi_j} \, \re^{{\rm i}(\phi_i-\phi_j)} \right] , 
\end{eqnarray}
for which the corresponding magnitudes $|\, \rho_{ij} \, |$ cannot in general be simplified, in the way we saw with \eqref{magnitudes for initial state}, on account of the possible correlations between the random probabilities and the random phases.
Indeed, the transformed probabilities $\{\pi_i \}_{i = 1,\,.\,.\,.\,,\,n}$ and phases 
$\{\phi_i \}_{i = 1,\,.\,.\,.\,,\,n}$ can take a variety of forms, depending on the 
precise nature of the system-environment interaction, and our goal is to give rather general conditions on these variables sufficient to give rise to decoherence. 

As a simple but nontrivial example, it will be useful first to consider the case of an
environment consisting of an apparatus that measures the value of 
the ${\hat X}$, without recording the outcome. The probability transformation takes the form 
\begin{eqnarray}
p_i \to \pi_i = \mathds 1 \{X = x_i\}, 
\end{eqnarray}  
where the random variable
$X$, which takes values in some set $\{x_i \}_{i = 1,\,.\,.\,.\,,\,n}$, denotes the outcome of the measurement. Here, for each value of $i$ we write
${\mathds 1}\{X=x_i\}$ 
for the indicator function that equals one if $X=x_i$ and zero otherwise. Note that $\pi_i$ is a random variable for each value of $i$. The random phases $\phi_i$ are equal to their initial constant values $\theta_i$ in this example and for the random state vector representing the system after its interaction with the measuring apparatus we obtain 
\begin{eqnarray}
|\Psi \rangle = \sum_{k=1}^n{\mathds 1}\{X=x_k\} \, \re^{{\rm i}\theta_k} |X_k\rangle  .
\end{eqnarray}  
The initial density matrix of the system has the matrix elements 
\eqref{initial matrix elements} in the ${\hat X}$-basis, whereas 
after the measurement the density matrix is diagonal in the ${\hat X}$-basis and takes the form 
\begin{eqnarray}
\hat \rho = \sum_{k=1}^n p_k |X_k \rangle  \langle X_k  |. 
\end{eqnarray} 

This example leads us on to the traditional ``information loss" interpretation for decoherence, for we see that the 
von Neumann entropy of the state of the system increases from the initial value 
\begin{eqnarray}
H^{\rm{vN}}_0 = - {\rm tr} \, (\hat r \log \hat r) = 0, 
\end{eqnarray} 
since $\hat r$ is pure, to a final value of
\begin{eqnarray}
H^{\rm{vN}}_f = - {\rm tr} \, (\hat \rho \log \hat \rho) =  -\sum_k p_k \log p_k,
\end{eqnarray} 
which is positive, hence corresponding to an information loss. 

Clearly, there are 
other maps of the form \eqref{random transformation} that can be used to model the system-environment interaction. Whatever 
form these take,  it remains that the von Neumann entropy increases, since the state \eqref{transformed density matrix} after the transformation is mixed, whereas the initial 
state is pure. 
The argument follows that the system must have somehow lost information since the von Neumann entropy of the state of the system has increased. The environment, on the 
other hand, so it is argued, has gained this information, for instance, by detection of the measurement outcome, even if this is not recorded in a corresponding transformation of the state of the system into a new pure state.  
The key difficulty with this train of thought leading to the claim that decoherence 
implies information loss is that the notion of decoherence is basis dependent, whereas 
the von Neumann entropy is basis independent. Hence, while decoherence is observed in the ${\hat X}$-basis, the information that is ``lost'' has no obvious relation to the decoherence basis, which is puzzling. 

With these considerations in mind, we hope to offer a clearer understanding of the role of information in the decoherence process based on the following points. 
First, we note that the allegedly  ``lost'' information, as measured by the increase in von Neumann entropy, is information lost by the
observer, not the system as such. Initially, the observer 
has knowledge of the state of the system, as a pure state, but owing to random interactions between system 
and environment, the information held by the observer is replaced by a 
statistical description of the state, represented by a density matrix. The observer loses track of the system as a pure state, of which they no longer have knowledge, which is replaced by knowledge of the ensemble average. 
On the other hand, the information 
about ${\hat X}$ implicit in the state is encoded in the Shannon entropy 
associated with the squared magnitudes of the expansion coefficients of the state in the ${\hat X}$-basis. We shall show that the difference between the initial Shannon entropy and the expected value of the
final Shannon entropy is positive. 
Thus, in this sense,  it is the system that gains information, not the environment, on average, in the process of decoherence. 

The key point here is that when we consider together the random pure state, the associated ensemble average, and the preferred basis, there are two different entropies that need to be taken into account. One reflects the knowledge of an external observer, and the other refers to information implicit in knowledge of the random pure state of the system. The latter is not accessible in the ordinary way to an external observer, though we can imagine a kind of ``demon'' who would have such access. Both entropies are relevant to the discussion and correspond to the different levels of information implicit in the various structures under consideration. 

It should be emphasized that entropy (whether it be the von Neumann entropy associated with a density matrix or the Shannon entropy associated with a pure state) is not an observable in the strict sense in quantum mechanics -- it represent rather a summary statement about the level of knowledge that can be possessed concerning a system and as such it may vary according to which aspect of the system is being considered. 
In this respect, the entropies that we consider have a status not unlike that of the Heisenberg uncertainty associated with an observable in a particular state. For although the uncertainty is not an ``observable" in the sense of Dirac, it can be given an operational meaning in terms of the statistics of measurements on ensembles. 

We support our conclusions concerning information gain and loss with four propositions. Proposition 1 asserts that if the information implicit in the state of a 
system concerning an observable ${\hat X}$ increases, this leads to decoherence of the density matrix of the system
in the basis of ${\hat X}$. In setting up the conditions under which this result holds we make use of a general framework for information gain based on the use of conditional expectations. In particular, we use the notion of information as it arises in Kolmogorov's set-theoretic characterization of conditional expectation.

Proposition 2 is a rather more general result -- namely, if a 
random transformation of the form (\ref{random transformation}) preserves the diagonal elements of the 
density matrix in the basis of ${\hat X}$, this leads to decoherence in that basis. This 
 is perhaps surprising -- since conventionally one 
associates decoherence with the 
decay of the off-diagonal elements -- so-called dephasing; whereas we show that \textit{there is a more fundamental 
characterization of decoherence purely in terms of the diagonal elements of the density 
matrix}. Thus, there is no need to examine the off-diagonal elements of the density matrix to determine whether the system has decohered. 

Proposition 3, which arises as a corollary to Proposition 2, shows that when there is decoherence, the variance of ${\hat X}$ 
on average decreases, reflecting the fact that decoherence implies information gain by the state of the system. 
Finally, in Proposition 4 we show, as promised, that the Shannon entropy of a system associated with the decoherence basis decreases, on average, on account of its interaction with the environment.
To avoid some trivial cases, we shall assume that the random variables $\{\pi_i\}$ are 
linearly independent, that is to say, that no set of constants $\{\lambda_i\} \in \mathbb R^n \backslash \{0\}$ exists such that 
\begin{eqnarray}\label{condition}
\sum_{k=1}^n \lambda_k\pi_k = 0.
\end{eqnarray} 
For instance, if $n=2$ then $\pi_1$ and $\pi_2$ are linearly dependent if and only if they are constant, because if \eqref{condition} holds there are two constraints $\pi_1 + \pi_2 = 1$ and 
$\lambda_1 \pi_1 + \lambda_2 \pi_2 = 0$ for some $\lambda_1, \lambda_2$, so  $\lambda_1 \neq \lambda_2$, so $ \pi_1 = \lambda_2 /( \lambda_2 - \lambda_1)$ and $ \pi_2 = \lambda_1 /( \lambda_1 - \lambda_2)$.

\vspace{0.2cm} 
\noindent \textbf{Proposition 1}. Suppose that the state of the system undergoes a 
random transformation of the form (\ref{random transformation}) in such a way that nontrivial information concerning the observable ${\hat X}$ is reflected in the new structure of the state. Then the density matrix of the system decoheres in the ${\hat X}$-basis. 

\vspace{0.2cm} 
\noindent \textit{Proof}. We begin with a few preliminaries. 
For $p \geq 1$ we say that a random variable $X$ lies in 
${\cal L}^p$  if ${\mathbb E}[|X|^p] <\infty$. Then its ${\cal L}^p$-norm is defined  by 
$\|X\|_p=({\mathbb E}[|X|^p])^{1/p}$. 
Then we have the Cauchy-Schwarz inequality, which in a probabilistic setting asserts 
that if $X, Y\in{\cal L}^2$ then $XY\in{\cal L}^1$ and  
\begin{eqnarray} \label{Schwarz inequality}
\left| \,{\mathbb E}[XY]\,\right|
 \leq {\mathbb E}\left[ \,|XY|\,\right] 
 \leq \|X\|_2 \, \|Y\|_2 \,, 
\end{eqnarray} 
with equality on the right if and only if $X$ and $Y$ are linearly dependent  \cite{Williams}. 

A rather natural way of characterizing the notion of information gain in many settings is by use of Kolmogorov's theory of conditional probabilities \cite{kolmogorov1933}. This theory
has the advantage of enabling one to work at a high level of generality. Specifically, 
let ${\mathscr I}$ denote any choice of an information set (viz., a sigma-algebra) that is not independent of the random variable $X$. What non-independence means is 
that the specification of ${\mathscr I}$ will provide nontrivial (albeit, generally, incomplete) information about the value 
of $X$. The corresponding informationally-rich random probabilities are given in terms of 
conditional expectations by
\begin{eqnarray}
\pi_i = {\mathbb P}( X=x_i \,|{\mathscr I}) = {\mathbb E}[ {\mathds 1}\{X=x_i\}\,|{\mathscr I}]  . 
\end{eqnarray} 
Then by the tower property of conditional expectation we have 
\begin{eqnarray} \label{crucial property}
{\mathbb E}[\pi_i] = {\mathbb E}[ {\mathbb E}[ {\mathds 1}\{X=x_i\} \,|{\mathscr I}]] = {\mathbb E}[ {\mathds 1}\{X=x_i\}] = p_i.
\end{eqnarray} 
Since the $\{\pi_i\}$ that are constructed in this way are centered at $\{p_i\}$ and linearly independent, and are such that $\sqrt{\pi_i}\in{\cal L}^2$ for each $i$, it follows that $\| \! \sqrt{\pi_i }\, \|_2 = \sqrt{p_i}$, and from \eqref{Schwarz inequality}  we have
\begin{eqnarray} 
{\mathbb E}\left[ \!\sqrt{\pi_i  \pi_j}\,\right] \, < \, \| \! \sqrt{\pi_i}\,\|_2 \,\, \|\!\sqrt{\pi_j}\,\|_2 
= \! \sqrt{p_i p_j} 
\label{inequalities} 
\end{eqnarray} 
for $i \neq j$. Here we have used the assumption that the $\{p_i\}$ are linearly independent to deduce a strict inequality. Next, we consider the magnitudes of the matrix elements \eqref{matrix elements of state} of the transformed state, given by
\begin{eqnarray} \label{magnitudes of matrix elements of state}
|\,\rho_{ij} \,|=  \left| \,{\mathbb E}\left[ \! \sqrt{\pi_i \,\pi_j} \, \re^{{\rm i}(\phi_i-\phi_j)} \right] \,\right| . 
\end{eqnarray}
If  $Z$ is any  complex-valued random variable with integrable real and imaginary parts, then 
\begin{eqnarray} \label{complex inequality}
{\mathbb E}[\,|\,Z\,|\,]\geq|\,{\mathbb E}[Z]\,|,
\end{eqnarray}
with equality only if $Z$ is constant. This follows by use of Jensen's inequality for a convex function of two real variables.
Since the argument of the expectation in  \eqref{magnitudes of matrix elements of state} is a complex random variable, we can use  \eqref{complex inequality} to
deduce that 
\begin{eqnarray} \label{inequality for matrix elements of state}
|\,\rho_{ij} \,| \leq   \,{\mathbb E}\left[ \,\left| \! \sqrt{\pi_i \,\pi_j} \, \re^{{\rm i}(\phi_i-\phi_j)}\right| \,\right] =
 {\mathbb E}\left[ \!\sqrt{\pi_i  \pi_j}\,\right] 
\end{eqnarray}
for $i \neq j$. Finally, if we combine  \eqref{magnitudes for initial state}, \eqref{inequalities}, and  \eqref{inequality for matrix elements of state}, we get
\begin{eqnarray} \label{basic inequality}
|\,\rho_{ij} \,| \leq  {\mathbb E}\left[ \!\sqrt{\pi_i  \pi_j}\,\right] < \! \sqrt{p_i p_j}  =  |\,r_{ij}\,| 
\end{eqnarray} 
for $i \neq j$ , which shows that the magnitudes of the off-diagonal matrix elements of the state after interaction with the environment will be strictly less than the corresponding magnitudes for the initial state, as claimed. \qed

If $\{\pi_i\}$ and $\{\phi_i\}$ are independent, then the leftmost inequality of \eqref{inequality for matrix elements of state} is an equality, but we still obtain \eqref{basic inequality} for $i \!\neq \! j$. Thus, if $\{\pi_i\}$ and $\{\phi_i\}$ are statistically dependent, the inequalities become even sharper. A special case of Proposition~1 was established in \cite{BT} where the
information set ${\mathscr I}$ is generated by a random variable correlated to 
$X$. Here we generalize that result to the case where the conditioning is with respect to any $\sigma$-algebra. Note that if ${\mathscr I} = \sigma[X]$, the case of maximal information, then the off-diagonal elements of the density matrix vanish. 
When the $\{\phi_i\}$ are random but the
$\{\pi_i\}$ are constant and hence equal to the $\{p_i\}$, the inequality (\ref{inequalities}) 
is an equality, even though there decoherence arises from the random 
phases. Such a case arises as a result of a random unitary transformation generated 
by the observable ${\hat X}$, as in the dynamics considered by Peres 
\cite{Peres} and Adler \cite{Adler}. Then decoherence leads neither to
information gain nor to information loss, owing to unitarity. 
It will be noted that a crucial consequence of assuming that $\{\pi_i\}$ is a conditional probability is equation \eqref{crucial property}. In fact, we can isolate that sole property and use it as a basis for the argument going forward, leading to the following. 

\vspace{0.2cm} 
\noindent \textbf{Proposition 2}. Suppose that the quantum state undergoes a transformation 
of the form (\ref{random transformation}) with the property that the probabilities are conserved  in the basis of 
an observable ${\hat X}$, so
\begin{eqnarray} 
{\mathbb E}[\pi_i]=p_i . 
\label{mean condition}
\end{eqnarray} 
Then we obtain \eqref{basic inequality} for all  $i \neq j$. 
\vspace{0.2cm} 

\noindent \textit{Proof}. Clearly $\sqrt{\pi_i } \in {\cal L}^2$. Further, by use of  \eqref{mean condition}  we have 
$\| \! \sqrt{\pi_i }\, \|_2 = \sqrt{p_i}$. Hence, by the Cauchy-Schwarz inequality (\ref{Schwarz inequality}), the assumed independence of the $\{p_i\}$, and the complex inequality \eqref{complex inequality}, we get \eqref{basic inequality}, as 
claimed. \qed

Though Propositions 1 and 2 appear similar, there is a 
difference. In Proposition~1, we assume that the gain in information is achieved by conditioning with respect to an information set, which 
 implies  \eqref{mean condition}. In 
Proposition~2 we have  
assumed the mean condition, without stipulating a particular mechanism for information gain. Therefore, if a quantum state undergoes a
random transformation such that the diagonal elements of the density matrix are conserved in the basis of ${\hat X}$, then the state will 
decohere in that basis.

\vspace{0.2cm} 
\noindent \textbf{Proposition 3}.  Let the quantum state undergo a 
transformation (\ref{random transformation}) such that \eqref{mean condition} holds. Then 
\begin{eqnarray} 
{\mathbb E}\left[ 
\langle \Psi|{\hat X}^2|\Psi\rangle - \langle \Psi|{\hat X}|\Psi\rangle^2 \right] 
< 
\langle \psi_0 |{\hat X}^2|\psi_0 \rangle - \langle \psi_0 |{\hat X}|\psi_0\rangle^2 , 
\label{uncertainty reduction} 
\end{eqnarray} 
showing that the averaged uncertainty of ${\hat X}$ is reduced when the system is in the state $|\Psi\rangle$. 
\vspace{0.2cm} 

\noindent \textit{Proof}. 
For $W \in {\cal L}^2$ we have ${\mathbb E}[W^2] \geq  \left( {\mathbb E}[W] 
\right)^2$ by Jensen's inequality, with equality if and only if $W$ is constant. 
Hence if we set $W = \sum_{k} \pi_k x_k$, which is square-integrable, we obtain 
\begin{eqnarray}
{\mathbb E} \left[\,\left(\sum_{k=1}^n \pi_k x_k \right)^2\right] > 
\left( {\mathbb E}\left[ \sum_{k=1}^n \pi_k x_k \right]\, \right)^2 = \left( \sum_{k=1}^n p_k x_k\right)^2 \!, 
\end{eqnarray} 
by use of the mean condition \eqref{mean condition} and our assumption that the $\{\pi_i\}$ are linearly independent. Changing the sign and adding equal terms to each side, we obtain
\begin{eqnarray}
{\mathbb E} \left [ \sum_{k=1}^n \pi_k x_k^2 \right] - {\mathbb E} \left[ 
\left(\sum_{k=1}^n \pi_k x_k \right)^2 \right] 
<  \sum_{k=1}^n p_k x_k^2 -\left( \sum_{k=1}^n p_k x_k \right)^2 \!, 
\end{eqnarray}
again by use of \eqref{mean condition}. But this gives (\ref{uncertainty reduction}), as 
claimed. \qed

In \cite{BT,DCB} it was conjectured that if a system decoheres in 
the basis of an observable ${\hat X}$, then the variance of ${\hat X}$ will on average 
decrease. This conjecture was shown in \cite{BT} to be true for two-level systems. 
Here we have generalized the result to systems of any size, with a 
simpler proof. 

Another way of characterizing the uncertainty of ${\hat X}$ in the state is in terms of the Shannon entropy. The initial value of the Shannon entropy of the system is 
\begin{eqnarray}
H^{\rm{Shan}}_0 = -\sum_k p_k \log p_k \, ,
\end{eqnarray} 
whereas for the final value in expectation we have
\begin{eqnarray}
H^{\rm{Shan}}_{\!f} = -{\mathbb E} \,\left [\sum_k \pi_k \log \pi_k \right] .
\end{eqnarray} 

\vspace{0.1cm} 
\noindent \textbf{Proposition 4}. Suppose that a quantum state undergoes a 
transformation of the form (\ref{random transformation}) such that \eqref{mean condition} holds. Then  
\begin{eqnarray}
{\mathbb E} \left [ H^{\rm{Shan}}_{\!f} \right] < H^{\rm{Shan}}_0.
\end{eqnarray} 

\noindent \textit{Proof}.
It is not difficult to check that the function $I : \mathbb R^+ \backslash \{0\} \to \mathbb R$ defined  by $I(x) = x \log x$ for $x>0$ is convex. 
Hence, for any strictly positive integrable random variable $W$ it holds that 
\begin{eqnarray}
{\mathbb E} [ W \log W ] \geq {\mathbb E} [W] \log {\mathbb E} [W] 
\end{eqnarray} 
by Jensen's inequality, with equality only if $W$ is constant. It follows by use of \eqref{mean condition} that 
\begin{eqnarray}
{\mathbb E} [ \pi_k \log \pi_k ] > {\mathbb E} [\pi_k ] \log {\mathbb E} [\pi_k ] = p_k \log p_k,
\end{eqnarray} 
for each $k = 1, . \, . \, . \,, n$, and hence 
\begin{eqnarray}
-\sum_k p_k \log p_k > -{\mathbb E} \,\left [\sum_k \pi_k \log \pi_k \right],
\end{eqnarray} 
as claimed.  \qed

Therefore, under any random transformation of the state that preserves the diagonal elements of the density matrix in the basis of a preferred observable ${\hat X}$, the ensemble average of the Shannon entropy of the system decreases from its initial value following an interaction with the environment.
Our results show that we can model the decoherence of a quantum system due to interactions with the environment by a random perturbation of the state of the system that in expectation conserves the probability law for the outcome of the measurement of a preferred observable. 
In fact, one sees now that we can take this property of conservation of probability to be a {\em definition} of what we mean by an interaction with the environment that generates decoherence of the dephasing type. 
Then decoherence in the basis of the preferred observable implies that the system has on average acquired information about that 
observable, as quantified by a decrease in the Shannon entropy. 

In modelling the state transition by a random perturbation, we step outside of the usual formalism of quantum mechanics with the inclusion of random elements. 
This can be justified on the grounds that we are in essence putting forward a {\em reduced-form model} for decoherence, rather than, say, proposing a fundamentally new dynamics for the state vector. 
One can take a similar view on the dynamics of continuous-time reduction models, since these also exhibit decoherence of the density matrix 
while preserving the mean condition
 \cite{Gisin1989, ABBH2001, BH2023}. 

In summary, we have shown that decoherence implies information gain by the quantum system, rather than loss, contrary to conventional wisdom, and that the decay of the off-diagonal elements of the density matrix follows from a conservation law that preserves the diagonal elements of the density matrix. Let us emphasize that our conclusions are to a large extent independent, at least qualitatively, of the precise choice of the random perturbation. That is, while an external observer will have no access to the details of the way in which the system has been perturbed, our results show that in whatever state the system has landed, an observer can be sure that the Shannon entropy has on average reduced. We hope that our findings will lead to new 
ways of understanding the emergence of classicality from an informational perspective.

\vspace{0.2cm}
\begin{footnotesize}
\noindent {\bf Acknowledgements}. The authors thank Simon Saunders 
for stimulating discussions. 
DCB acknowledges support from EPSRC 
(EP/X019926) and the John Templeton Foundation (grant 62210). 
The views expressed in this publication are those of 
the authors and do not necessarily reflect the views of the 
John Templeton Foundation.
\end{footnotesize}

\end{document}